\documentclass[aps,prb,twocolumn,superscriptaddress,amsmath,amssymb,nofootinbib]{revtex4-2}
\usepackage{graphicx}  
\usepackage{xcolor}
\usepackage{dcolumn}   
\usepackage{bm}        
\usepackage{verbatim}   
\usepackage{gensymb}
\usepackage{epstopdf}
\usepackage{footmisc}
\usepackage{natbib}

\usepackage{color}
\usepackage{multirow}
\usepackage{physics}
\usepackage{braket}
\usepackage{mathtools}
\usepackage{hyperref}
\usepackage{lineno}

\newcommand{\MNS}{Mn$_{1/3}$NbS$_2$}
\newcommand{\MfNS}{Mn$_{1/4}$NbS$_2$}
\newcommand{\CNS}{Cr$_{1/3}$NbS$_2$}
\newcommand{\CNSe}{Cr$_{1/3}$NbSe$_2$}
\newcommand{\Mn}{{$^{55}$Mn}}
\newcommand{\Cr}{{$^{53}$Cr}}
\newcommand{\Nb}{{$^{93}$Nb}}

\newcommand{\bhf}{$B_{\mathrm{hf}}$}

\newcommand{\Bhf}{$\mathbf B_{\mathrm{hf}}$}
\newcommand{\TN}{$T_{\mathrm{m}}$}
\newcommand{\muB}{$\mu_{\mathrm{B}}$}

\begin{document}

\title{ Disorder-resilient transition of Helical to Conical ground states in M$_{1/3}$NbS$_2$, M=Cr,Mn}

\author{M. Sahoo}
\affiliation{ Dipartimento di Scienze Matematiche, Fisiche ed Informatiche, Universit\`a di Parma, Parco Area delle Scienze 7A, I-43100 Parma, Italy}
\affiliation{Leibniz IFW Dresden, Helmholtzstra\ss{}e 20, D-01069 Dresden, Germany}
\affiliation{Institut f\"ur Festk\"orper- und Materialphysik and W\"urzburg-Dresden Cluster of Excellence ct.qmat, Technische
Universit\"at Dresden, 01062 Dresden, Germany}

\author{P. Bonf\`a}
\affiliation{ Dipartimento di Scienze Matematiche, Fisiche ed Informatiche, Universit\`a di Parma, Parco Area delle Scienze 7A, I-43100 Parma, Italy}

\author{A. E. Hall}
\affiliation {Department of Physics, University of Warwick, Coventry, CV4 7AL, United Kingdom}

\author{D. A. Mayoh}
\affiliation {Department of Physics, University of Warwick, Coventry, CV4 7AL, United Kingdom}

\author{L.~T.~Corredor}
\affiliation{Leibniz IFW Dresden, Helmholtzstra\ss{}e 20, D-01069 Dresden, Germany} 

\author{A. U. B Wolter }
\affiliation{Leibniz IFW Dresden, Helmholtzstra\ss{}e 20, D-01069 Dresden, Germany}

\author{B.~Büchner}
\affiliation{Leibniz IFW Dresden, Helmholtzstra\ss{}e 20, D-01069 Dresden, Germany}
\affiliation{Institut f\"ur Festk\"orper- und Materialphysik and W\"urzburg-Dresden Cluster of Excellence ct.qmat, Technische
Universit\"at Dresden, 01062 Dresden, Germany}

\author{G. Balakrishnan}
\affiliation{Department of Physics, University of Warwick, Coventry, CV4 7AL, United Kingdom}

\author{R.~De Renzi}
\email[Corresponding address: ]{roberto.derenzi@unipr.it}
\affiliation{ Dipartimento di Scienze Matematiche, Fisiche ed Informatiche, Universit\`a di Parma, Parco Area delle Scienze 7A, I-43100 Parma, Italy}

\author{G.~Allodi}
\affiliation{ Dipartimento di Scienze Matematiche, Fisiche ed Informatiche, Universit\`a di Parma, Parco Area delle Scienze 7A, I-43100 Parma, Italy}

\date{\today}
\begin{abstract}
The discovery of chiral helical magnetism (CHM) in \CNS~and the stabilization of a chiral soliton lattice (CSL) has attracted considerable interest in view of their potential technological applications. However, there is an ongoing debate regarding whether the sister compound, \MNS, which shares the same crystal structure, exhibits similar nontrivial properties which likewise rely on the lack of inversion symmetry at the magnetic ion.  
In this study, we conduct a comprehensive investigation of the magnetically ordered states of both compounds, using  \Cr, \Mn~and \Nb~nuclear magnetic resonance.  Our results, supported by density functional calculations, detect in a high-quality single crystal of \CNS~all the signatures of the monoaxial CHM in a magnetic field, identifying it as a reference case for NMR. The detailed understanding of this prototypic behavior provides a reference for \MNS.
Despite the much larger density of specific defects in this second material, we confirm the presence of a CHM phase in the Mn compound, characterized by a very large critical field for the forced ferromagnetic phase ($\approx$ 5 T for the applied field along $c$).

\end{abstract}
\maketitle


\section {Introduction}
 Layered transition metal dichalcogenides (TMDCs), denoted as $XY_2$, where X represents a transition metal (e.g., Nb, Ta) and Y represents a chalcogen (e.g., S, Se), have garnered significant attention owing to their intriguing electrical and optical properties \cite{choi2017recent,manzeli20172d,wei2018various,krasnok2018nanophotonics,mak2016photonics}. Despite being nonmagnetic, the weak van der Waals bonding between the layers in $XY_2$ allows for remarkable tunability within 2D limits and it also enables the intercalation of magnetic 3d transition metals, leading to the formation of ternary TMDC-based magnetic compounds (M$_{1/3}$XY$_2$, where M can be Cr, Mn, V, Fe, Co, Ni) exhibiting diverse structural and magnetic properties \cite{parkin19803,friend1977electrical},  including strong spin-orbit coupling.
Sulfur-based compounds, in particular, feature localized moments arising from the spins of unpaired electrons at the intercalant sites because the orbital moment is often quenched due to the crystal field effects from neighboring sulfur atoms \cite{mito2019observation, friend1977electrical}. The magnetic properties are largely dictated by the type and concentration of the intercalants; for instance, Cr, and Mn-based TMDCs exhibit dominant ferromagnetic interactions, while for V, Co and Ni-based ones the dominant exchange is antiferromagnetic \cite{parkin19803,Hall.PhysRevB.103.174431}. Recent studies on M$_{1/3}$NbS$_2$, where M can be Cr or Mn, have unveiled the emergence of chiral helimagnetism \cite{kousaka2009chiral,kousaka2016long,togawa2012chiral,braam2015magnetic,dai2019critical,karna2019consequences,Mayoh2022}. This discovery opens up new avenues for exploring novel properties and potential technological applications, marking an exciting development in the field of TMDCs \cite{nair2020electrical,okumura2019spin}.

The chiral magnetism observed in \CNS, which belongs to the non-centrosymmetric space group P6$_{3}$22 (no. 182), originates from the arrangement of intercalant atoms forming a $\sqrt{3} \times \sqrt{3}$ superlattice in a bilayer triangular configuration with full occupation of the 2c Wyckoff site.
However, this superlattice 
is the natural candidate for site occupation disorder (SOD) by Cr atoms, which is realized by a small occupation of 2b and 2d sites and a sub-unity occupation of the 2c site.
\MNS{} is also reported to crystallize in the same space group but with a more sizable partial occupation of the Mn 2b and 2d sites. In this compound a small fraction of few percent of \MfNS{} has also been reported. \cite{hall2022comparative}

The broken inversion symmetry, described by the Dzyaloshinskii-Moriya (DM) interaction, competes with the in-plane ferromagnetic Heisenberg exchange interaction (J), leading to the emergence of chiral helimagnetic order (CHM) along the c-axis at zero magnetic fields. The origin of the ferromagnetic interaction is attributed to the RKKY mechanism \cite{rahman2022rkky} and to Hund's exchange \cite{Sirica.CommunPhys.3.65}. The direction of spin rotation and the chirality of the system are determined by the sign of the DM vector \cite{boswell1978ordering,hulliger1970magnetic,togawa2012chiral,braam2015magnetic}. The ratio between the DM vector and J gives the periodicity of the CHM phase in the sample;
however, the chirality of these compounds may be compromised by structural disorder \cite{dyadkin2015structural}. 
Under the influence of a magnetic field, the CHM transforms into various nonlinear magnetic structures. For instance, applying the field along the chiral axis ($\mathbf H\parallel \hat c$) results in a chiral conical phase (CCP), whereas applying the field within the $ab$ plane ($\mathbf H \perp \hat c$) leads to the formation of a chiral soliton lattice (CSL) phase. Both phases arise due to different magnetic energies competing with the incommensurate CHM phase. Further increasing the magnetic field transforms the incommensurate phase into a commensurate forced ferromagnetic (FFM) phase above a certain critical magnetic field, aligning all the spins along the field direction \cite{boswell1978ordering,hulliger1970magnetic,togawa2012chiral,braam2015magnetic,aoki2019anomalous}. 
The reported critical field values for both compounds and each phase vary slightly, possibly due to the itinerant character of magnetism \cite{karna2019consequences}. For $\mathbf H \perp \hat c$, in our \CNS~sample the CSL phase has been observed until 40 mT, and after that, the FFM phase emerges below $T_C$ =111 K \cite{hall2022comparative}, while  for $\mathbf H\parallel \hat c$ the CCP and the FFM phases arise above 0.5 and 1.5 T, respectively \cite{bykov2023magnetic}. Simultaneously, for the \MNS~compound, only the FFM phase (or saturation) is confirmed between 25 to 80 mT along $\mathbf H \perp \hat c$ below  $T_C$ =45 K and 3.5T to 4 T along $\mathbf H\parallel \hat c$ direction \cite{hall2022comparative,karna2019consequences,dai2019critical}.


While the magnetic phase diagram of \CNS is well established, the isostructural compound \MNS~ remains a topic of discussion. It has been suggested to exhibit a similar chiral helimagnetic (CHM) and topological soliton lattice \cite{kousaka2009chiral,karna2019consequences,karna2021annihilation}. However, recent reports on this compound have not been able to distinguish between trivial ferromagnetic domains with 
$\pi$ domain walls and non-trivial ferromagnetic regions periodically divided by chiral soliton kinks of spins ($2\pi$ domain walls)~\cite{hall2022comparative,karna2021annihilation}. The observation of a large pitch length in the helimagnetic phase and the presence of short-range ferromagnetic domains of $\approx 250$ nm, as noted by Karna et al.~\cite{karna2019consequences}, despite not excluding the CSL, complicate the identification.

In addition to magnetometry and more direct identification offered by neutron scattering and electron microscopy, various spin structures and magnetic phases can be explored using local microscopic techniques such as muon spin relaxation ($\mu$SR) and nuclear magnetic resonance (NMR). In particular $\mu$SR has helped to identify different phases by analyzing the amplitude of the magnetic fraction in both zero field and applied fields, suggesting the presence of a helimagnetic phase below 50~K in \CNS~\cite{braam2015magnetic}. Similarly, zero-field NMR has been employed to try to identify the various valence states resulting from disorder in the compound \cite{ogloblichev2017magnetic,ogloblichev2018valence}.

\begin{figure}[ht]
    \includegraphics[width=.95\columnwidth]{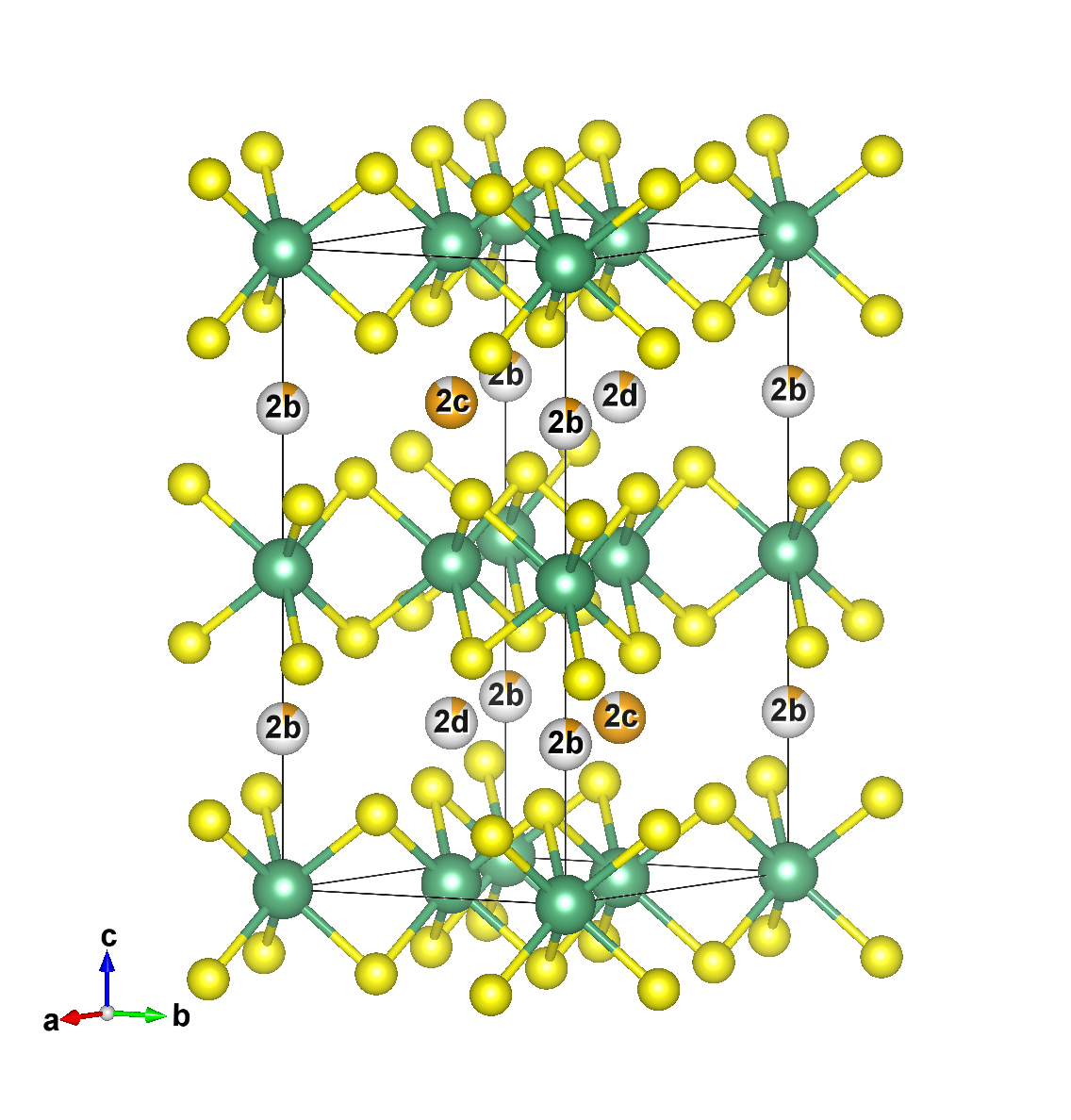}
    \caption{Layer structure and qualitative occupancies of the TM atoms (orange balls, with their Wyckoff notation, yellow balls are S atoms and green balls are Nb.}
 \label{fig:TM}
\end{figure}
A special role in these intercalated dichalcogenides is played by the partial occupancy of the transition metal (TM) intercalation layer: the TM occupies one of three possible sites in a hexagonal pattern (see Fig.~\ref{fig:TM}), $2b$ ,$2c$ and $2d$ in Wyckoff notation, of which ideally only $2c$ is occupied. However, in real crystals two types of defects may be expected: a non vanishing occupancy of $2b,2d$ sites, known as site-occupancy disorder (SOD), and, possibly, stacking faults along the $c$ axis, due to the weak Van der Waals bonds between NbS$_2$ layers in the pristine dichalcogenide. Both defects produce disorder, giving rise to a distribution of random local structures around any given atom, depending on the relative positions of the defect(s). \MNS\ is more prone to have these disorders compared to \CNS. In the present paper, we employ NMR, which might help to detect this disorder via e.g. an inhomogeneous line broadening of its resonance spectra and probe at the same time the mean thermodynamic properties of the material. Thus, NMR is a powerful probe to resolve disorder and its implications on both the local and average magnetic structure.


The presentation is organized as follows. The experimental details are described in \ref{sec:exp}.  In Sec.~\ref{sec:commonalities} we describe the common expected spectral patterns and in Sec.~\ref{sec:Cr.CNS} we present  an extensive set of zero-field (ZF) and field-dependent \Cr~NMR measurements on our \CNS\ single crystals in two different field orientations, which identify the different field ranges of the CSL, CCP and FFM phases, characteristic of a monoaxial chiral helimagnet \cite{kousaka2009chiral}, setting a reference standard for the still controversial case of the \MNS\ compound.  Moreover, we directly observe that the phase transition in \CNS\ from chiral helimagnetic (CHM) to paramagnetic is of the first order. Additional insight is obtained from \Nb~NMR in Sec.~\ref{sec:Nb.CNS}. 
The \Mn~NMR results from the \MNS~spectra and the assignment  supported by DFT calculations are presented in Sec.~\ref{sec:MNS}. 
The unique attribution of the spectral features is complicated by an occupancy of the Mn $2b$ and $2d$ Wyckoff sites, ideally empty, notably larger than that of Cr in the nearly ideal \CNS, resulting in a multi-modal inhomogeneous magnetic broadening. Despite this shortfall, the demonstration that \MNS~is also a chiral helical magnet is confirmed by the observation of large amplitude spin echo oscillations that reveal the same type of interplay of quadrupolar and magnetic interactions as in the \CNS~ case. Discussion with conclusions follows in Sec.~\ref{sec:discussion}.


\section{Experimental Details}
\label{sec:exp}
\subsection{Samples}
\label{sec:samples}
Polycrystalline and single crystal samples of \CNS~and \MNS~ studied in this work were synthesized by solid-state reaction and chemical vapor transport techniques, respectively, using the procedures described in \cite{hall2022comparative}, where their structural, magnetic characterization, and phase diagram along $\mathbf H \perp \hat c$ is extensively reported. Additional magnetization data on the Mn crystal are shown in Fig.~S4 in the Supplemental Material \cite{SM} (which also includes references \cite{hall2022comparative, elk, PhysRevLett.77.3865, PhysRevB.40.3616, 5.0005082,QE-2017,QE-2009}). Single crystals are in the form of platelets, approximately 3 mm along the longest edge. The phase purity of the polycrystalline materials was checked using X-ray powder diffraction. \cite{SM} Single-crystal x-ray diffraction was used to ascertain the non-centrosymmetric space group. Laue x-ray back-reflection studies were carried out to establish the crystalline quality as well as to check the crystallographic orientations of the obtained single crystals. \cite{hall2022comparative}
\subsection {NMR experiments}
\label{Sec:NMR}

The NMR spectra were acquired utilizing the HyReSpect home-built phase coherent broadband spectrometer \cite{allodi2005hyrespect} equipped with a 9~T magnet and a He-flow sample insert. The very broad spectra were recorded point by point at each frequency, in a Hahn spin-echo refocusing pulse sequence,  P--$\tau$--P. The P pulse duration and intensity were optimized for maximum signal, with the shortest $\tau$ delay, constrained by the apparatus dead time of a few microseconds. Spectra were reconstructed from the spin-echo Fourier transform amplitude at each frequency, adjusted for frequency-dependent sensitivity, and the nuclear Boltzmann factor, corrected for the $T_2^{-1}$ relaxation rates obtained by fitting the same spin-echo amplitude vs. delay ($2\tau$) to a relaxation function (exponential decay or Eq.~\ref{eq:spinecho}). Peaks in the spectra were assigned to each quadrupole transition (\Cr), or each hyperfine component (\Mn), here generically labeled $\alpha$. The best fit of each peak often requires more than one Gaussian component,
\begin{equation} \label{eq:Gaussian}
G_{\alpha}(\nu)= \sum_{i=1}^n\frac {A_{\alpha,i}}{\sqrt{2\pi}\sigma_{\alpha,i}}\,\exp\left[-\frac 1 2\left( \frac {\nu-\nu_{\alpha,i}}{\sigma_{\alpha,i}}\right)^2\right],
\end{equation}
whose mean frequency was computed as its first-moment $\overline{\nu_\alpha}$ by the weights $w_{\alpha,i}$:
\begin{equation} 
\label{eq:first-moment}
\overline{\nu_\alpha} = \sum_i w_{\alpha,i}\nu_{\alpha,i},\quad
w_{\alpha,i}=\frac {A_{\alpha,i}}{\sum_k A_{\alpha_k}}.
\end{equation}

The ZF and very low field nuclear spin echoes were measured using a non-resonant probe circuit under a substantial radio-frequency (rf) enhancement $\eta_{\mathrm{rf}}=H_1^*/H_1$, where $H_1$ is the applied field at the rf $\omega$, and $H_1^*$ is the response, the induced $\omega$ oscillating component of the large hyperfine field. \cite{sidorenko2006mn,meny2021nuclear}. The enhancement is progressively suppressed by the increasing applied fields, requiring the use of a tunable resonant probe.

\subsection{NMR and NQR}
\label{sec:NMR-NQR}

In the ordered magnetic phase, the full spin Hamiltonian for the nuclear spin $\mathbf{I}$ is \cite{Abragam}
\begin{align}
\label{eq:full-Hamiltonian}
    {\cal H} &= {\cal H}_{\mathrm {Z}} + {\cal H}_{\mathrm {Q}} \nonumber\\
    &= - h\gamma (\mu_0 \mathbf H + \mathbf{B}_{\mathrm {hf}})\cdot\mathbf I + \frac {h\nu_{\mathrm Q}}{3V_{zz}}\mathbf I \cdot \mathbf V \cdot \mathbf I,
\end{align}
\begin{equation}
    \label{eq:nuQ}
    \nu_Q = \frac{3eV_{zz} Q}{2I(2I-1)h}.
\end{equation}
where $\gamma$ is the nuclear gyromagnetic ratio, $\mu_0$ is the vacuum permeability constant, $\mathbf{V}$ is the electric field gradient (EFG) tensor and $Q$ is the quadrupolar coupling constant.
For both compounds and all three nuclei considered here the hierarchy is always ${\cal H}_{\mathrm {Z}}> {\cal H}_{\mathrm {Q}}$, hence the quantization axis is that of the hyperfine field $\mathbf B_{\mathrm{hf}}$, which may be rotated by the application of $\mathbf H$.  

We assume that the coupling tensors in the ideal structure are nearly cylindrical, with principal axis $z$ along $\hat c$, the EFG tensor asymmetry parameter $\eta = (V_{xx}-V_{yy})/V_{zz}\approx 0$ and $B_{\mathrm{hf}}^x\approx B_{\mathrm{hf}}^y$, in agreement with both the best fits and the DFT predictions (Sec.~\ref{sec:DFT}) and  justifying Eqs.~\ref{eq:frequencyab}-\ref{eq:Deltanu} below.

The conventional description of the static hyperfine field is
\begin{equation}
    \label{eq:hyperfine}
    \mathbf {B}_{\mathrm {hf}} = \mathbf {\cal{A}}\cdot\mathbf S,
\end{equation}
where $\mathbf{\cal{A}}$ is the effective hyperfine coupling dominated by the nearly isotropic on-site term for \Cr~and \Mn~NMR. Additional isotropic transferred couplings to nearest neighbor (nn) 3d ions are included in $\mathbf {\cal A}$, justifying a distribution of isotropic hyperfine tensor values when the local nn occupancy is different from that predicted by the crystal group. This effect is  negligible for the Cr spectra, relevant for Nb and even more for Mn, as discussed in Sec.~\ref{sec:discusMNS}.

\section{NMR Results}
\label{sec:NMRresults}

\begin{figure*} 
    \includegraphics[width=0.8\textwidth]{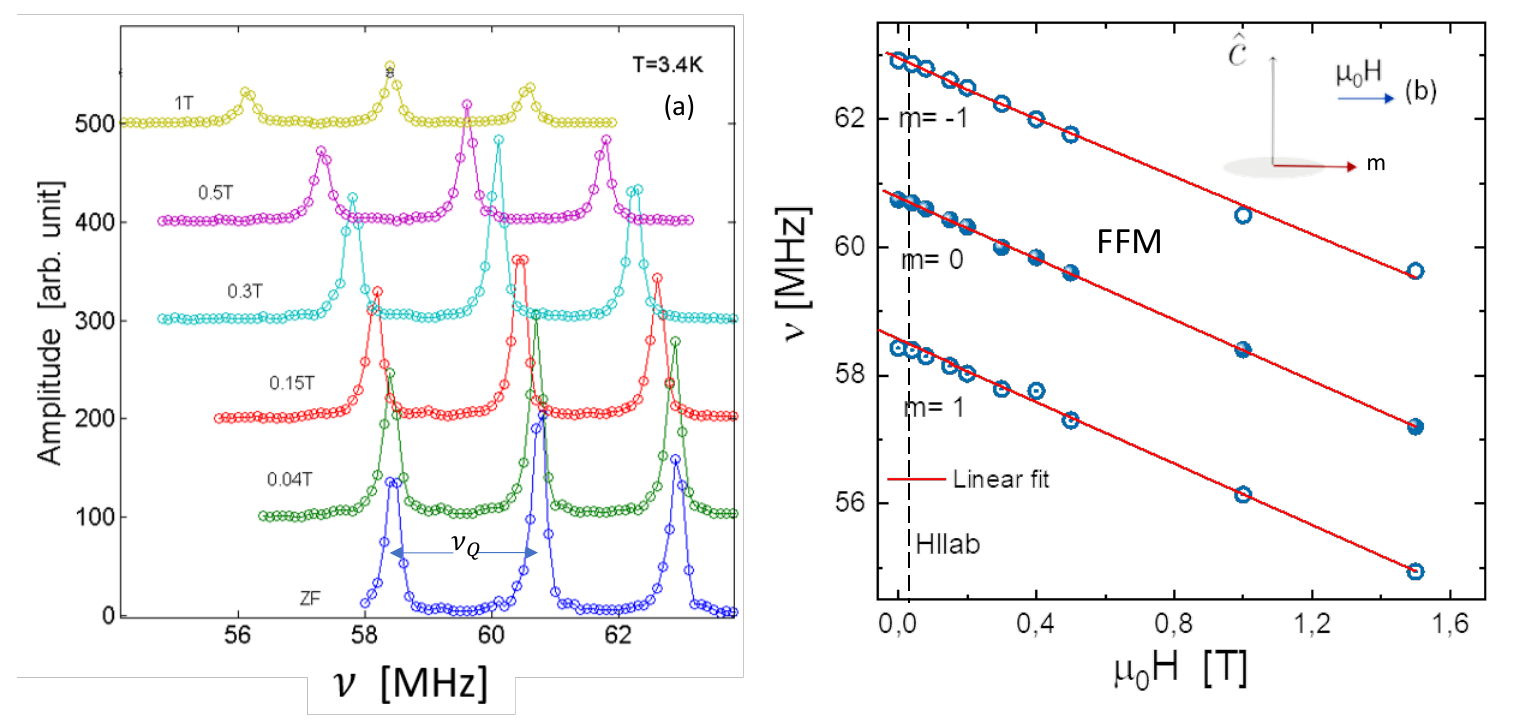}
    \caption{Representative \Cr~NMR spectra at $T=3.4$ K for (a)$\mathbf H \perp \hat c$, displaced vertically for clarity, with best fit; (b) the corresponding best fit peak frequencies vs. H with their three linear regressions (red solid lines)} 
 \label{fig:Cr_NMR}
\end{figure*}
In this section we describe how NMR results, despite being largely distinguished by different relative scales of the interactions appearing in the spin Hamiltonian and the different sample characteristics, eventually demonstrate that the magnetic properties of \CNS~and \MNS~ are very similar.


\subsection {Commonalities of the NMR spectra.}
\label{sec:commonalities}
The properties of the \Cr, \Mn~and \Nb~ nuclear isotopes are summarized in Tab.~\ref{tab:nuclei}.
\begin{table}[!ht]
    \caption{Nuclear spin $I$, gyromagnetic ratios $\gamma$, quadrupole moments $Q$ and natural abundance of \Cr,\Mn,\Nb} 
    \label{tab:nuclei}
    \begingroup
    \setlength{\tabcolsep}{7pt} 
    \renewcommand{\arraystretch}{1.1} 
    \begin{tabular}{ l |  c | c | c | c }
    \hline
    \hline
        & $I$ & $\gamma$ & $Q$ & Natural
        \\
        Isotope& & [MHz/T] & [fm$^2$] & abundance
        \\
        \hline
        \Cr & $\frac 3 2$ & 2.4115 & -15(5) \cite{Ertmer1982}  &  0.095 \\
  &&&-8.4 \cite{Stephenson2016}&    \\  
        \Mn & $\frac 5 2 $ & 10.5763 &  33(1)  & 1.0 \\
        \Nb & $\frac 9 2 $ & 10.4523 &  32(2) & 1.0 \\
        \hline
\end{tabular}
\endgroup
\end{table}
The single-crystal low-temperature NMR spectrum from each of them is qualitatively understood considering that the principal component of the nearly isotropic hyperfine field tensor $|{\cal A}| \approx {\cal A}_x\approx\ {\cal A}_z$  dominates on the smaller quadrupole coupling ($\nu_{\mathrm Q}\ll\nu_{\mathrm {hf}} \approx \gamma | \mathbf{B}_\mathrm{hf}| $). This produces a pattern of $2I $ frequencies, in first approximation equally spaced and centered at the Larmor frequency corresponding to the internal field at the nucleus. Since the hyperfine tensor and EFG tensor that couples to the nuclear electric quadrupole moment share the same principal axis, $\hat z = \hat c $, the two simple experimental conditions are with $\mathbf H$ either parallel or perpendicular to $\hat c$. In the first case, the angle $\theta=\theta(H)$ between the electron spin $\mathbf S$ and the easy-plane ($\perp \hat c$) has a non-trivial dependence on $H$. In the second case $\theta(H)=0$.
We can define
\begin{equation}
    \mathbf{S} = S \left(  \cos \theta \hat{x} + \sin \theta \hat{z} \right) \, .
\end{equation}
By combining the hyperfine field (Eq.~\ref{eq:hyperfine}) and the applied field,  the local field in the two conditions are \footnote{The relative sign of $H$~and the hyperfine field is due to the electron moment being opposite to the electron spin.}
\begin{align}
    B_{loc}^\perp &= A_x S - \mu_0 H \\
    B_{loc}^\parallel &= \sqrt{ S^2 A_x ^2 \cos^2\theta + (S{\cal A}_z\sin\theta-\mu_0 H)^2}.
\end{align}

The resulting frequency patterns in first-order perturbation are
\begin{equation}
    \nu_m^\perp(H) = \gamma B_{loc}^\perp   + \Delta\nu_m(0), \label{eq:frequencyab}
\end{equation}
 and 
\begin{equation}
    \nu_m^\parallel(H,\theta) = \gamma B_{\mathrm{loc}}^\parallel +\Delta \nu_m(\theta), \label{eq:frequencyc}
\end{equation}
where the transition label is $-\frac {2I-1} 2\le m\le \frac{2I-1} 2$ and
\begin{equation}
\Delta\nu_m(\theta) =  m \nu_{\mathrm Q} \frac {3\sin^2\theta -1}2. \label{eq:Deltanu}
\end{equation} 
Eqs. \ref{eq:frequencyab}-\ref{eq:Deltanu} predict $2I =3$, 5, and 9 lines for Cr, Mn and Nb, respectively and the field dependence is close to the exact one, but best fits of the data are refined by numerical diagonalization of the full spin Hamiltonian \ref{eq:full-Hamiltonian}, beyond first-order approximation. 

\subsection{\Cr~NMR in \CNS}
\label{sec:Cr.CNS}

Figure \ref{fig:Cr_NMR} (a) shows the expected triplet of NMR lines obtained in zero applied field (ZF) for the Cr single crystal, together with the shifted spectra measured with $\mathbf H\perp \hat c $ at 3.4 K. The ZF central frequency is around 61 MHz (i.e $|$\bhf$| = 25.34$ T).  
Triplet spectra are fitted to a minimal set of narrow Gaussian components.  The peak position for each of the three components is defined as its center of gravity (the first moment, see Eq.~\ref{eq:first-moment}). The rigid shift of these triplet peaks is towards lower frequencies, shown in Fig.~\ref{fig:Cr_NMR} (b) for this crystal orientation, with a slope equal to ${^{53}\gamma}$, indicating that the applied and internal field add collinearly according to Eq.~\ref{eq:frequencyab} in the forced ferromagnetic (FFM) phase expected for this easy plane chiral magnet. \cite{kishine2015} 

The origin of the small detected Gaussian broadening in the single crystal spectra is the very small degree of SOD, which produces different configurations of few distant Cr moments occupying the 2b and 2d defective Wyckoff sites. Similar, but broader spectra are detected in the polycrystalline sample  \cite{SM}, witnessing its larger SOD.

\begin{figure*}[ht]
    \includegraphics[width=1\textwidth]{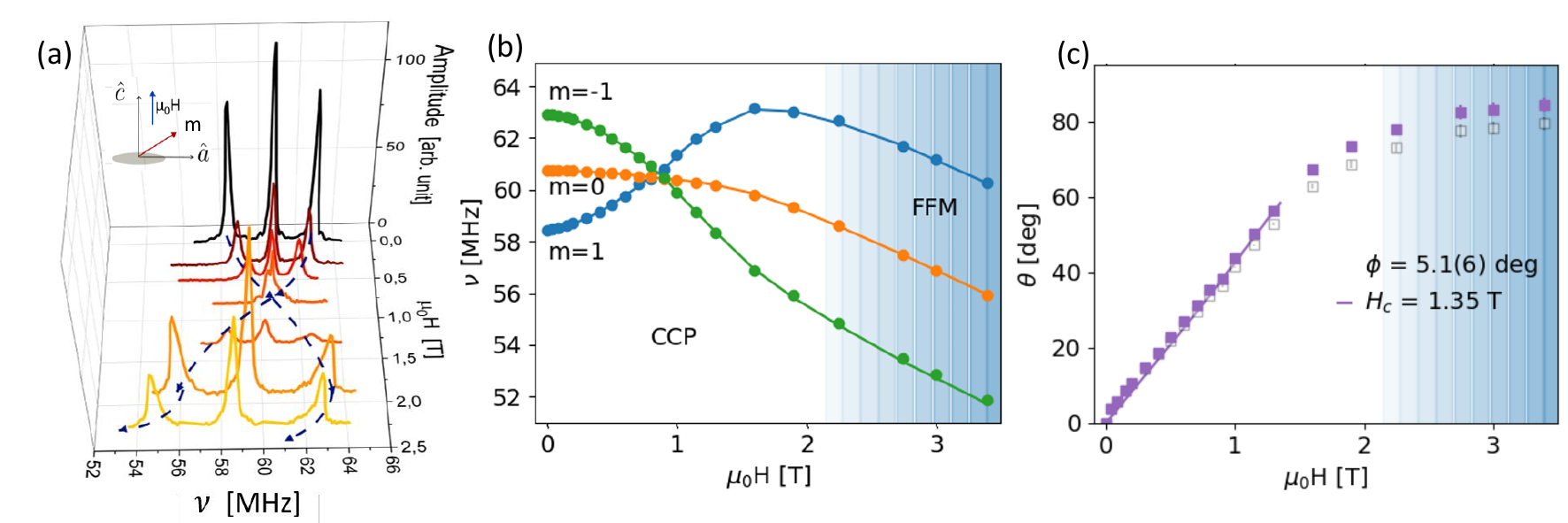}
    \caption{(a) Representative \Cr{} NMR spectra for $\mathbf H \parallel \hat c$ at $T=3.4$~K. (b)\Cr~peak frequencies vs.~field from the $\mathbf{H}\parallel \hat c$ spectra shown in (a) with best-fit curves (solid lines, see text).  (c) Field dependence of the best fit conical angles $\theta(H)$ with (purple) and without (gray) correction for experimental misalignment $\phi$ between  the field direction and $\hat c$ (symbols), with best fit to Eq.~\ref{eq:meanfieldc}.}
 \label{fig:conical}
\end{figure*}

The more complex pattern predicted by Eq.~\ref{eq:frequencyc} for the other crystal orientation, $\mathbf H\parallel \hat c $ is displayed in Fig.~\ref{fig:conical} (a). Notably, the triplet splitting vanishes at the angle $\theta=\sin^{-1}\sqrt{\frac 1 3}$, in agreement with Eq.~\ref{eq:Deltanu}. The peak positions are shown in Fig.~\ref{fig:conical} (b) vs.~ field. In first order, they agree with Eq.~\ref{eq:frequencyc}, but the global best fit (solid line) is obtained by numerical diagonalization, optimizing the common hyperfine and quadrupolar parameters reported in Tab.~\ref{tab:hfq}, together with a separate, local angle $\theta(H)$ for the three data points at each experimental field value. The latter are displayed in Fig.~\ref{fig:conical} (c) and, taking carefully into account the small experimental misalignment, \cite{SM} they agree  with mean field predictions \cite{kishine2015,SM}
\begin {equation}
\label{eq:meanfieldc}
\theta(H) = \sin^{-1}\left(\frac H {H_c}\right)
\end{equation}
(bear in mind that $\theta$ is the field angle from the plane). From the fit we obtain $\mu_0 H_c = 2S(\sqrt{J^2+D^2}-J+K_\perp)/g\mu_B = 1.35(1)$ T, $g$ is the Cr $S=3/2$ Land\'e factor, $J,D,K_\perp$ are the Heisenberg, Dzyaloshinskii-Moriya exchange and the easy plane anisotropy constants, respectively. 
It should be noted that a small demagnetization field is also present and is not included in our analysis of the local angle $\theta(H)$. Its value depends on the shape of the crystal and requires numerical computation. An upper estimate for this contribution is $\mu_0 M_0 \sim 0.18$~T which is less than 15\% of the estimated critical field.

\begin{figure}[ht]
    \includegraphics[width=0.8\linewidth]{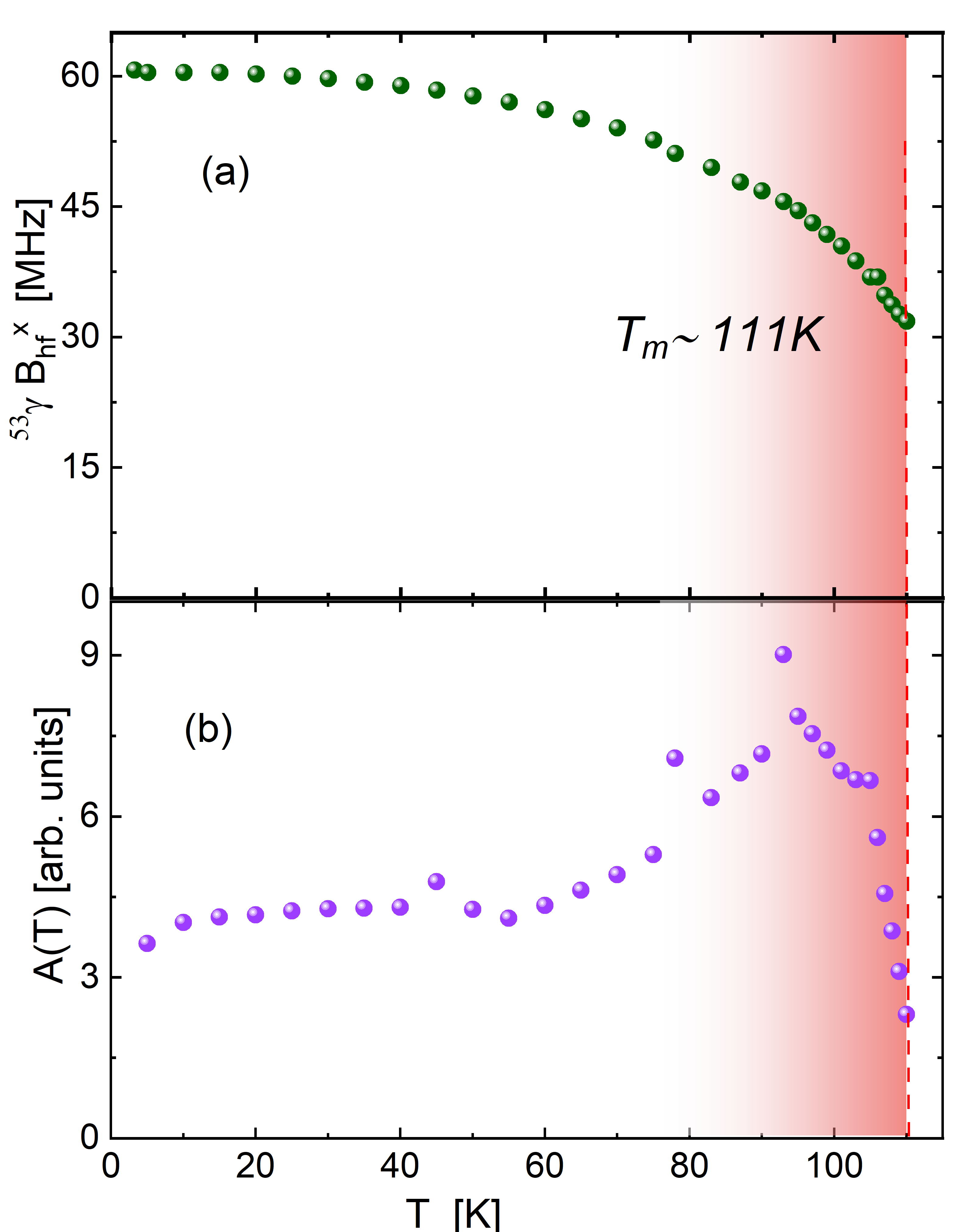}
    \caption{Temperature dependence of: (a) the \Cr~hyperfine frequency ${^{53}\gamma} B_{\mathrm{hf}}^x \propto\mu(T)$; (b) its NMR amplitude $A(T)$, corrected for the nuclear Curie law at zero external field.}
 \label{fig:Cr_temp}
\end{figure}

Let us turn now to the microscopic magnetization, measured by the hyperfine field (Sec. \ref{sec:NMR-NQR}), and largely dominated by the contact on-site contribution. 
The typical value for the contact hyperfine field in transition metal ions is 
$\approx 10$ T/$\mu_{\mathrm B}$ \cite{WatsonFreeman61, Shaham}. 
Therefore, the mean hyperfine field value of $\sim$26 T in Tab.~\ref{tab:hfq} roughly corresponds to $2.6\, \mu_{\mathrm B}$. According to Eqs.~\ref{eq:frequencyab} and \ref{eq:Deltanu}, the temperature dependence of the central transition frequency $\nu_0$  in the ZF \Cr-NMR triplets 
is proportional to the atomic moment $\mu(T) $ on Cr, and leads to the values plotted in Fig.~\ref{fig:Cr_temp} (a). Notice that the decrease of the NMR amplitude $A(T)$ in Fig.~\ref{fig:Cr_temp} (b), in the highlighted pre-transition region, 90 K $<T\le$ \TN, is typical of a first-order transition. It shows that paramagnetic domains start developing well below the transition temperature \TN, and that the average electron spin value $S(T)$ in the residual ordered domains remains large (the hyperfine field at \TN~is still more than half of its zero temperature value).\footnote{Instead, the smoother increase of $A(T)$ from 60 to 90 K is due to magnetic anisotropy reduction, which provides a larger ferromagnetic enhancement see Sec.\ref{sec:exp}; notice that the first-order loss of amplitude is clearly visible, {\em despite} this amplified NMR sensitivity.} 

\begin{table}[!ht]
    \caption{\Cr~global best fit parameters for Fig.~\ref{fig:conical}; $\phi$ accounts for a possible misalignment.} 
    \label{tab:hfq}
    \begingroup
    \setlength{\tabcolsep}{7pt} 
    \renewcommand{\arraystretch}{1.1} 
    \begin{tabular}{ l |  c | c | c }
    \hline
    \hline
        $\nu_{\mathrm{Q}}$ & ${^{53}{\cal A}}_{x}S$ & ${^{53}{\cal A}}_{z}$S & $\phi$
        \\
        $[\mbox{MHz}]$  & [T] & [T] & [deg]\\
        \hline
        4.49(3)&  25.172(3) & 26.63(1)  &  5.1((6)\\
        \hline
     \end{tabular}
     \endgroup
\end{table}

\subsection{\Nb~NMR in \CNS.}
\label{sec:Nb.CNS}
 
\begin{figure}[ht]
    \centering
    \includegraphics[width=0.48\textwidth]{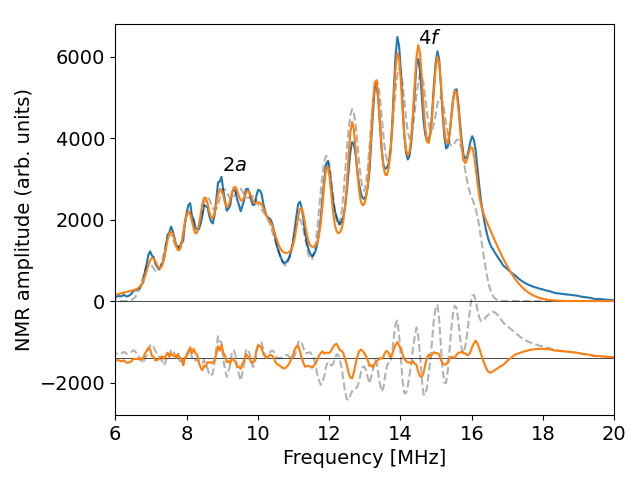}
    \caption{Zero-field (blue curve) \Nb~NMR spectrum at $T=1.3$ K in \CNS~with its 2-site (gray dashed) and 4-site (orange solid) best fits, see \cite{SM}. The residues are plotted with a negative offset and the same colors.}
    \label{fig:NbZF}
\end{figure}
The ZF NMR spectrum of \Nb, is best obtained on the much larger polycrystalline sample due to the improved signal to noise ratio at lower frequencies. Unfortunately this sample displays a somewhat larger degree of SOD. \cite{SM} The blue curve in Fig.~\ref{fig:NbZF} shows two replica of the nine quadrupolar transitions predicted by Eq.~\ref{eq:frequencyab} 
for nuclear spin $I=\frac 9 2$, that can be attributed to the two inequivalent sites Nb $2a$ and $4f$  in space group 182, in the multiplicity ratio 1:2. The separation of the two nonuplets is proportional to the $T=0$ difference  ${(^{93}{B}}_{\text{hf}, 4f}-{^{93}{B}}_{\text{hf},2a})\approx 0.5$ T of their ZF hyperfine fields, roughly 0.8 and 1.3 T, respectively. These fields are too large to be due to distant dipoles on Cr ions (a rough estimate of the dipolar field yields 0.2 T). The larger field value indicates a dominant contribution from a local moment on Nb, due to Cr-Nb hybridization. 

The best fit curves in Fig.~\ref{fig:NbZF} are obtained by centering Gaussians at the quadrupole perturbed NMR transitions, similar to Eq.~\ref{eq:frequencyab}, but calculated up to second order (see \cite{SM}). They require pairs of Nb sites in the amplitude ratio 1:2, as per Wyckoff sites $2a$ and $4f$. The simplest model, with just these two main sites (Tab.~\ref{tab:Nbfit}) corresponds to the gray dashed curve. A refined model with an additional pair of sites corresponding to 6\% of the Nb nuclei  is shown by the orange curve.
\begin{table}[]
    \centering
     \caption{\Nb~quadrupolar and hyperfine parameter of the 2-site best fit (in SM) in Fig.~\ref{fig:NbZF}) .}
    \label{tab:Nbfit}
    \begin{tabular}{c | c | c | c | c}
    \hline
    \hline
        Wyckoff & fraction & $\nu_{\mathrm{Q}}$ & $\nu_{\mathrm{hf}}$ & $\delta\nu$\\
         site & & [MHz] & [MHz] & [MHz] \\
         \hline
        $2a$ & 0.33 & 0.9 & 8.9 & 0.19 \\
        $4f$ & 0.67 & 1.23 & 13.8 & 0.21\\
    \end{tabular}
   
\end{table}

\begin{table}[]
    \centering
     \caption{\Nb~quadrupolar and hyperfine parameter of the 4-site best fit (in SM) shown in Fig.~\ref{fig:NbZF}).}
    \label{tab:Nbfit-4sites}
    \begin{tabular}{c | c | c | c | c}
    \hline
    \hline
        Wyckoff & pair & $\nu_{\mathrm{Q}}$ & $\nu_{\mathrm{hf}}$ & $\delta\nu$\\
         site & fraction & [MHz] & [MHz] & [MHz] \\
         \hline
        $4f_1$ & \multirow{2}{*}{0.94} & 0.83 & 8.82 & 0.15 \\
        $2a_1$ & & 1.2 & 13.8 & 0.11\\
        $4f_2$ & \multirow{2}{*}{0.06} & 1.8 & 10.7 & 0.74 \\
        $2a_2$ & & 0.9 & 14.9 & 0.65\\
    \end{tabular}
   
\end{table}

In the 4-site model the parameters of the majority pair are very similar to those of the simpler 2-site fit. The second pair has the same fixed amplitude ratio and broader satellite widths therefore hinting at a contribution coming from sites close to site occupancy defects in the Cr plane.
Each Nb is coupled to several, say $n$, Cr in occupied 2c sites and shares a potential coupling path to several unoccupied 2b, 2d sites. Therefore a tiny fraction $x$ of SOD leading to unoccupied 2c or occupied 2b, 2d sites will alter the hyperfine coupling of $\approx nx$ Nb sites. A few percent of any of these defects are the likely source of the 6\% contribution in the 4 site fit.


\begin{figure}[htb]
    \centering
    \includegraphics[width=0.45\textwidth]{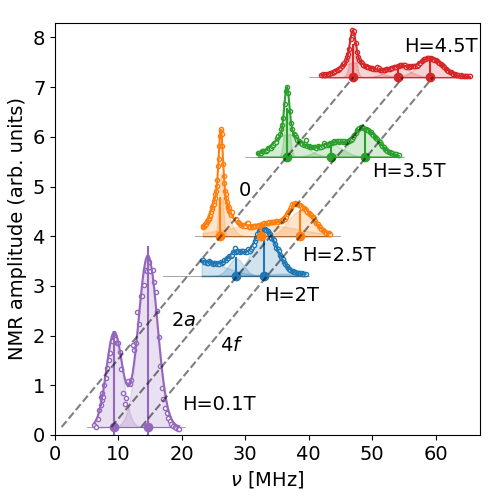}
    \caption{\Nb~NMR spectra in applied fields $H$ at $T=1.3$~K, shifted along the vertical axis proportionally to $H$. The lines correspond to ${^{93}\gamma}\mu_0H+\nu_{\mathrm{hf}}$.}
    \label{fig:Nb-vs-field}
\end{figure}

Figure \ref{fig:Nb-vs-field} shows that the application of an external field to the polycrystalline sample produces two or three broad peaks, from inequivalent Nb sites, labeled $\alpha=0,2a,4f$ (the second two by their Wyckoff notation). The large powder broadening prevents the resolution of the quadrupole pattern and the best fit of the 2a, 4f sites is obtained with a Gaussian each, in the appropriate fixed amplitude ratio. All peaks shift to higher frequencies following the corresponding dashed ${^{93}\gamma} \mu_o H + \nu_{\mathrm{hf}}^\alpha$ lines. The positive shift with field, opposite to that of \Cr~, indicates that the moment on Nb is antiparallel to that of neighbor Cr ions. The NMR data also evidences the sizable presence of Nb sites ($\alpha=0$) with a vanishing hyperfine field, fitted by two further Gaussians sharing the same center. Its origin is possibly due to vacancies in the $2c$ site which would lead to a negligible hybridization for the two $nn$ Nb nuclei, therefore strongly reducing their hyperfine field. Unfortunately, a more detailed analysis is not possible with the present powder data.


\subsection{\Mn~NMR in \MNS.}
\label{sec:MNS}

\begin{figure*}[ht]
    \includegraphics[width=0.8\textwidth]{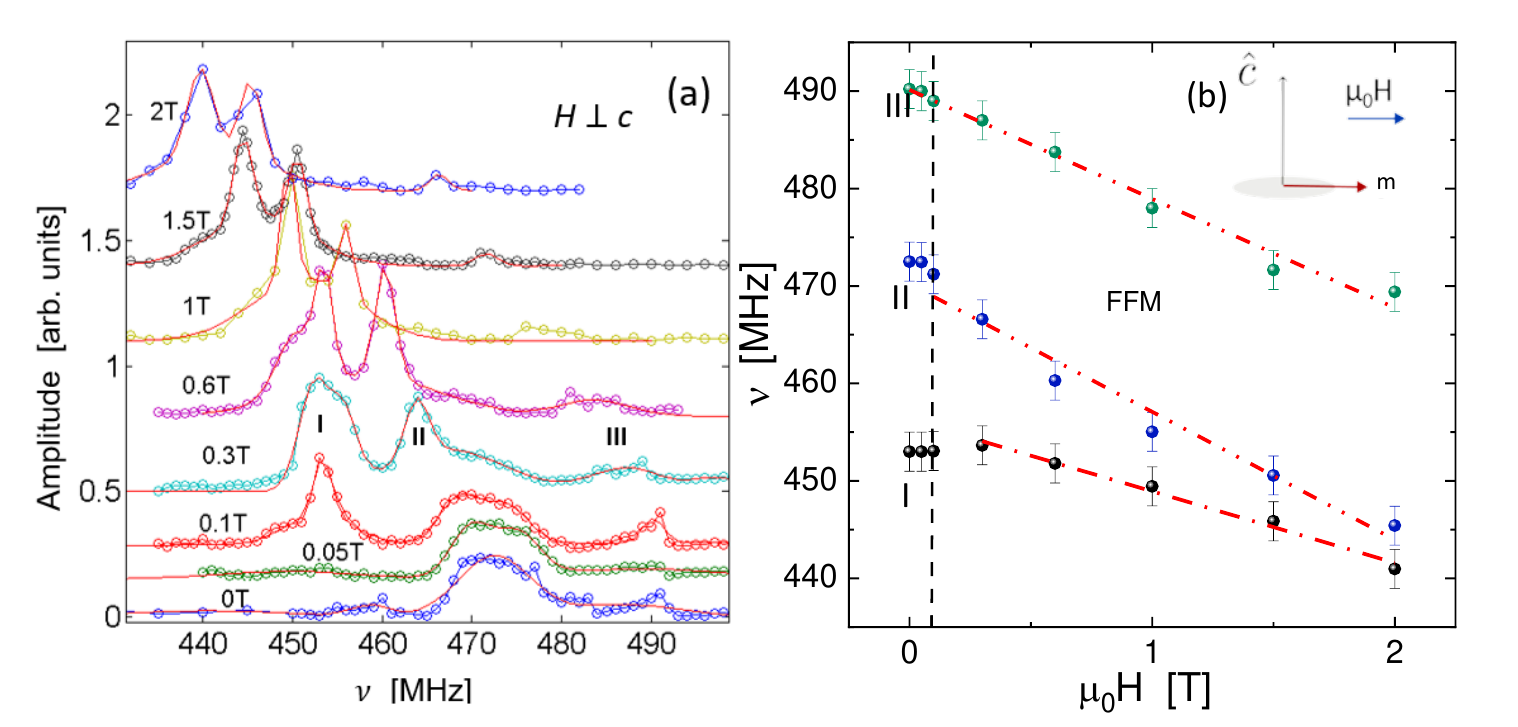}
    \caption{\MNS~single crystal: \Mn~NMR in zero and applied fields: (a) spectra for $\mathbf H \perp \hat c$; (b) field dependence of the frequency center of gravity, for $\mathbf H \perp \hat c$ (full symbols) and their three linear regressions (red dotted lines, just intended  as a guide to the eye). }
 \label{fig:Mn_hllab}
\end{figure*}

Figure \ref{fig:Mn_hllab} (a) and (b) show a selection of \Mn~NMR low temperature spectra obtained in a \MNS~single crystal at $T=1.7$ K from ZF up to $\mu_0 H=2$ T, with $\mathbf H \perp \hat c$ 
covering the range of 440–490 MHz. In low fields, they are centered around 470 MHz. We identify a few prominent components, consisting of overlapping peaks, which gradually change shape while shifting towards lower frequencies with the field. We may identify the first moment of each component (see Sec.~\ref{Sec:NMR} for details).
\begin{figure*}[ht]
    \includegraphics[width=1\textwidth]{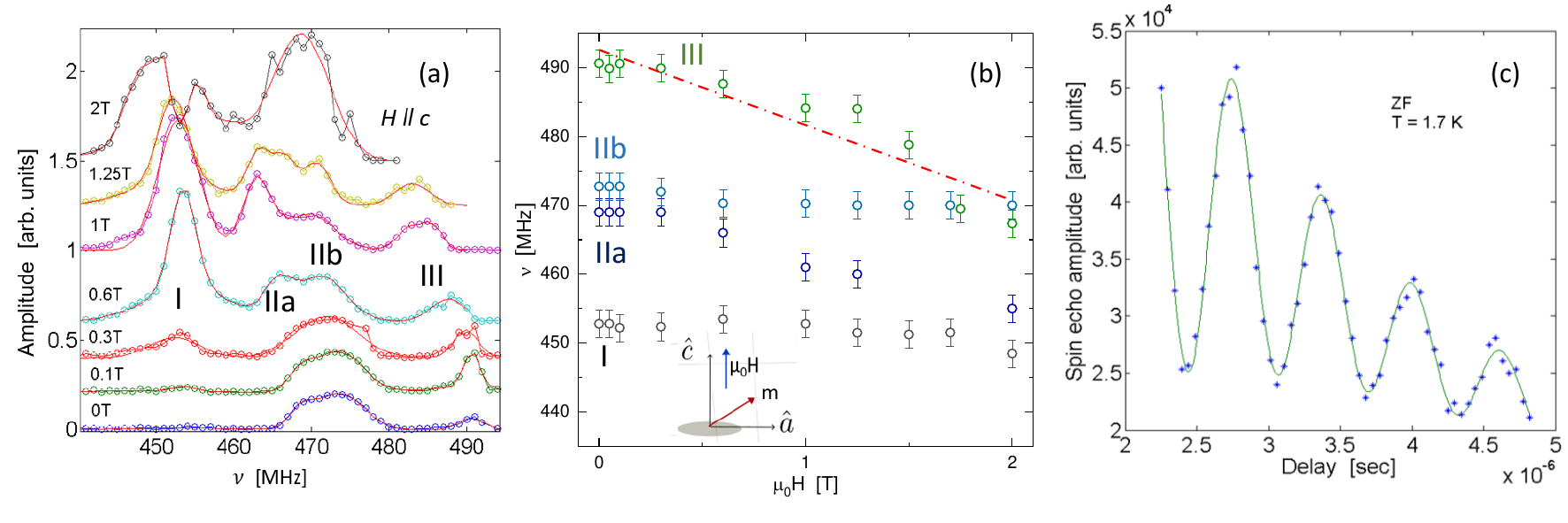}
    \caption{\MNS~single crystal: \Mn~NMR in zero and applied fields (a) spectra for $\mathbf H \parallel \hat c$; (b) field dependence of the frequency center of gravity, for $\mathbf H \parallel \hat c$ (open symbols and its linear linear regression, red dotted line ); (c) spin echo amplitude vs. pulse delay at $T=1.7$ K, $\mu_0 H=0$ T, 472 MHz displaying oscillations, with solid line best fit to  Eq.~\ref{eq:spinecho}.}
 \label{fig:Mn_hllc}
\end{figure*}

We must note a few experimental differences between the \Mn~case and the reference \Cr~data of Fig.~\ref{fig:Cr_NMR} (a) and Fig.~\ref{fig:conical}(a), before making a comparison. First of all, the \Mn~NMR amplitude at low fields is affected by very fast, field-dependent $T_2^{-1}$ relaxation rates, which, e.g., heavily suppress the 450 MHz component in ZF.\footnote{When $T_2$ is comparable to the dead-time of our NMR probe, the signal becomes marginal and difficult to reliably correct for relaxation, leading to a partial, so-called, \emph{wipeout}.} Furthermore, the much larger gyromagnetic ratio of \Mn~(roughly a factor four, see Tab.~\ref{tab:nuclei}) and the larger Mn moment per ion (roughly a factor two) modify the prediction of Eqs.~\ref{eq:frequencyab}, \ref{eq:frequencyc}, producing eight times larger hyperfine frequencies, in agreement with the experimental ratio of about 470 MHz to 60 MHz. This implies that magnetic defects should also yield larger inhomogeneous linewidths, scaling from 0.3 MHz on \Cr~ to 2.4 MHz on \Mn. However, the distribution of magnetic defects in \MNS~makes the spectral peaks even broader than this prediction, with linewidths well in excess of 5 MHz, which completely hides the quadrupolar splittings in this sample. 

If \MNS~were another textbook case, like  \CNS, a {\em single broad} line would be expected, with a field dependence similar to that of the central ${^{53}\nu_0}$ (Fig.~\ref{fig:Cr_NMR} (b), filled blue symbols, and Fig.~\ref{fig:conical} (b), orange symbols). Unfortunately, the \Mn~spectra show more than one broad component, and they are separated by up to 35 MHz, a much larger and more diverse splitting than that expected for the quadrupole pattern, which is composed of five much closer, equally spaced lines for \Mn, spin $I=5/2$. These quintuplets are convoluted by the larger magnetic inhomogeneous broadening as discussed above, hence hidden within the linewidth.  In conclusion, the origin of the multimodal distributions of hyperfine fields must stem from magnetically different local surroundings.  

Let us concentrate on panel (a) in Fig. \ref{fig:Mn_hllab}, showing the results for $\mathbf H \perp \hat c$. We have already mentioned three main components: the more intense peaks at lower frequencies, originating at 455 and 475 MHz, labeled I and II, respectively, and a much weaker peak around 490 MHz, labeled III in Fig.~\ref{fig:Mn_hllab} (a). We attribute components I and II to the majority \MNS~structure, in two distinct, possibly defect-free and defect-related configurations, further discussed in Sec.~\ref{sec:discusMNS}. 
The higher frequency of peak III indicates a fractionally larger valence, closer to Mn$^{2+}$ \cite{rahman2022rkky}, and a comparably larger magnetic moment. Despite the large error bars we can compare the field dependence of their first moments in Fig.~\ref{fig:Mn_hllab} (b)  with those of \Cr: for $\mathbf H \perp \hat c$ (green solid symbols), peak III shifts to lower frequency with a slope in agreement with ${^{55}\gamma}$ (dash-dotted red line). This is expected in an easy plane ferromagnet when the spin aligns along $\mathbf H$  and the magnetic moment (hence the hyperfine field) aligns opposite to $\mathbf H$. Peak III, of similar amplitude in both orientations, thus identifies a small \MfNS~fraction already observed in magnetization with a higher ordering temperature around 100 K  in samples of the same batch \cite{hall2022comparative}.  

Figure \ref{fig:Mn_hllab} (b) further shows that the first moments of components I and II for $\mathbf H \perp \hat c$ (solid blue, black symbols)  follow at high field a  behavior similar to that of III (green symbols), in agreement with Eq.~\ref{eq:frequencyab}, $m=0$, within the reduced accuracy granted by the large error bars. Only at lower fields, where a chiral soliton phase is expected, they depart from this linear field dependence, characterized by the full gyromagnetic ratio.
In contrast, the results for $\mathbf H \parallel \hat c$ , Fig.~\ref{fig:Mn_hllc} (b), reveal that the field dependence  of peak III (green symbols) is unaffected by field orientation, while the other peaks (open light and
dark blue symbols) have a much weaker field dependence in the whole field range, in qualitative agreement with the conical phase described  by Eq.~\ref{eq:frequencyc}, $m=0$ and demonstrated by the \Cr~$m=0$ transition of Fig.~\ref{fig:conical} (b). 

We obtain strong support for this tentative assignment by considering the peculiar spin-echo ($T_2$) relaxation of peak II, shown in Fig.~\ref{fig:Mn_hllc} (c), displaying large amplitude oscillations with a relaxing envelope. This behavior deviates considerably from the normal simple exponential echo decay. The oscillations are observed in ZF and in moderate fields in both peaks I and II of \MNS.  \cite{SM}. This is a well known phenomenon, \cite{voNidda2010} that arises, in the present context, from coherent beatings of pairs of quadrupole lines originating from the same individual nuclear spin.
The coherent oscillations in the spin echo relaxation amplitude vs. delay time $2\tau$, \cite{abe1966spin,Lombardi1996,voNidda2010}, (Fig.~\ref{fig:Mn_hllc}c and Fig.~S.1) are seen when irradiating a broad spectrum, due to an inhomogeneous distribution of interactions (i.e., changing from nucleus to nucleus), in the presence of a smaller, more coherent interaction (i.e. the same for all nuclei). The coherent interaction determines the same frequency splitting $\Delta\nu$ on each nucleus and many spectral isochromats come in pairs. An  rf spectral width broader than this splitting produces the beating of all pairs, showing up as an oscillation in the echo amplitude across the spectrum. A general expression is
\begin{equation}
\label{eq:spinecho}
    A(\tau) = [A_0 + A_1\cos(4\pi\Delta\nu \tau) + \phi)]\exp\left(-\frac {2\tau}{T_2}\right).
\end{equation}
Here, the inhomogeneous broadening is magnetic and the quadrupole separation $\Delta\nu_1(H)$ of Eq.~\ref{eq:Deltanu}, which depends on the spin angle $\theta=\theta(H)$, acts as the coherent splitting ($\phi$ is a generic phase).
\begin{figure}[ht]
    \includegraphics[width=0.8\linewidth]{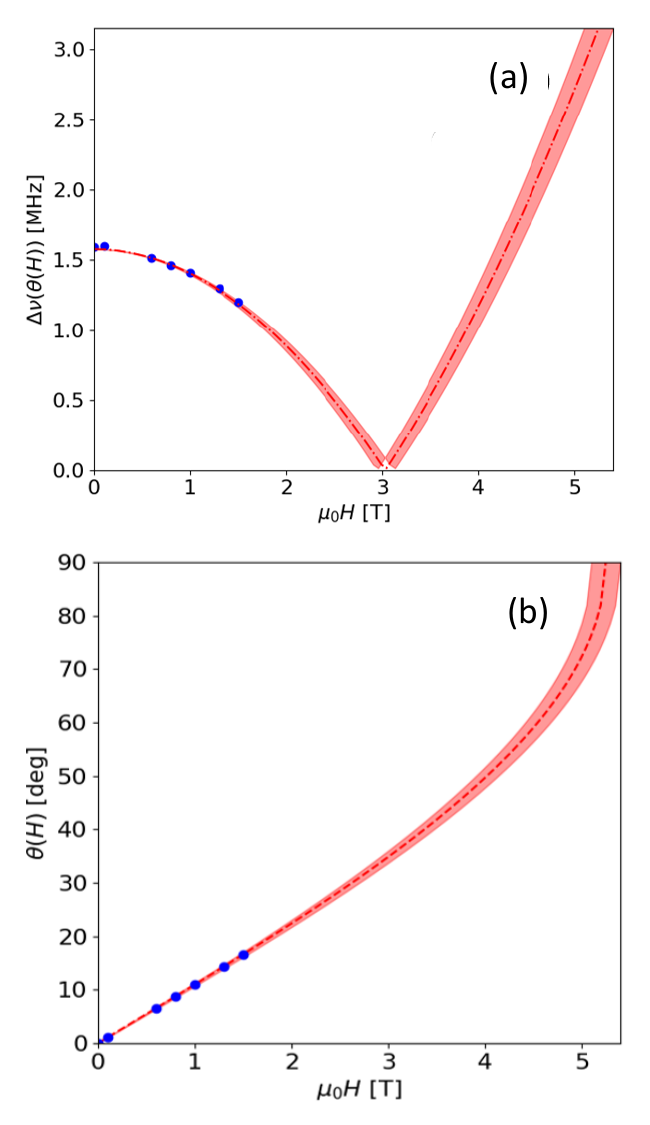}
    \caption{(a) Quadrupole splitting $|\Delta\nu(\theta(H))|$ of  \Mn~($\mathbf H || \hat c$) and (b) conical angle $\theta(H)$  vs. field, from spin-echo oscillation frequency, curves and red bands represent best fit to the absolute value of Eq.~\ref{eq:Deltanu} with Eq.~ \ref{eq:meanfieldc} and their standard deviation, respectively.}
 \label{fig:Mn_qd}
\end{figure}

We obtain the field dependent splitting plotted in Fig.~\ref{fig:Mn_qd} (a), for $H\parallel \hat c$ from the fit of the corresponding $T_2(H)$ spin-echo oscillations.  The field dependence can be further fitted assuming the monoaxial helical chiral model of Ref.~\cite{kishine2015} and Eq.~\ref{eq:meanfieldc}, obtaining the $\theta(H)$ values plotted in Fig.~\ref{fig:Mn_qd} (b) together with the corresponding mean field prediction. The good quality of both fits confirms at the same time the attribution of the lower frequency majority components to \MNS~and the validity of the conical model ($\mathbf H \parallel \hat c$) for this sample as well. The best-fit parameters are a quadrupole frequency $\nu_Q = 3.15(5)$ MHz and a critical field $\mu_0 H_c=5.0(5)$ T for the FFM phase which coincides with the saturation field measured in isothermal magnetization within a reasonable error bar, shown in Fig. S4 \cite{SM}. Misalignment uncertainties \cite{SM} are expected to be the same as for the Cr crystal, of the order of 5 degrees, and are included in the error estimate. Notice that $T_2$ oscillations are lost well before reaching this field value.

\subsection{DFT calculations}
\label{sec:DFT}
Ab initio simulations have been performed with a full potential formalism in order to determine both the EFG and the hyperfine field at the Cr and Mn nuclei.
The Elk code has been used to perform all simulations and the details are provided in the SM~\cite{SM}.
The experimental lattice structure has been used for all simulations and a ferromagnetic ground state approximates the long-wavelength helical order.
The resulting ground state magnetic moment is $3 \mu_{\text B}/\text{Cr}$ and $4.2 \mu_{\text B}/\text{Mn}$ for \CNS~and \MNS~respectively. Details are given in the SM~\cite{SM}. 

\subsubsection{DFT in the {\rm Cr} compound}
The EFG has the principal axis for the largest component aligned with the $c$ lattice parameter, and  $\eta=0$ for Cr. We obtain $V_{zz}=2.2 \times 10^{21}$~V/m$^2$ that results in a predicted quadrupole frequency (defined in Eq.~\ref{eq:nuQ}) $\nu_Q = 4.0$ MHz that compares very well with the experimental estimate provided in Tab.~\ref{tab:hfq} and agrees with previous reports \cite{Agzamova2022}.
The dominant contribution to the hyperfine field is of contact origin and its estimated value is $\approx 11$~T. The pseudo-dipolar and orbital contributions are about one order of magnitude smaller and are close to the accuracy limit of the present computational analysis. 
The final prediction, which includes contact, orbital, and dipolar interactions, results in 14 T, thus underestimating the experimental value of Tab.~\ref{tab:hfq}. We briefly mention that including $3s$ states in the valence generates a contact field of approximately 26 T. However, this prediction is expected to be less accurate, as detailed in the next section and in the SM.

\subsubsection{DFT in the {\rm Mn} compound}

Also in this case the largest component of the EFG tensor 
is aligned with $c$ and we obtained $V_{zz}=2.9 \times 10^{21}$~V/m$^2$, which results in $\nu_Q= 3.5$ MHz, which compares well with the experimental $\nu_Q=2|\Delta \nu_1(0)|=3.15(5)$ MHz.
The hyperfine parameters are dominated by the contact part also in this case, which is found to be 20 times larger than orbital and dipolar contributions, which are therefore neglected in the following discussion. 
The estimate for the contact field depends on the inclusion of relativistic effects for $3s$ electrons. When included in the core, their wave-function is obtained from the solution of the spin-polarized Dirac equation. This produces an estimate of about 30 T for the contact part, while keeping $3s$ in valence raises this value to 50 T. 
The former value is more reliable, but we mention that removing 3s semi-core states from valence also represent an approximation.


\section{Discussion and conclusions}
\label{sec:discussion}
\subsection{\CNS}
{\label{sec:discusCSN}}


Let us begin with the textbook Cr single crystal sample, where we observe a \Cr~~quadrupole triplet pattern, previously obscured in polycrystalline data \cite{ogloblichev2018valence}, likely due to lower sample quality. This observation is consistent with more recent \CNSe~~results \cite{ogloblichev2017magnetic,ogloblichev2021electronic}, supporting the conclusion that our crystal is of high quality and belongs to the $P6_{3}22$ space group.

The 
hyperfine field yields an estimate 
of the Cr magnetic moment of 2.6 \muB~at $T=0, H=0$, with a 13\% reduction over the spin-only Cr$^{3+}$ value of 3 \muB. Although older magnetization data \cite{Miyadai1983} report a value of 2.9 \muB~at 2 T, our result is in line with the observed \TN$=111$~K and critical field value $\mu_0H_c=1.35(1)$ T in our crystal, Fig.~\ref{fig:conical} (c). 

A small non vanishing moment on Nb of the order of 0.1-0.15 \muB, opposite to that of Cr is demonstrated by the \Nb\ hyperfine field, Fig.~\ref{fig:NbZF}, and the field dependence of the spectra, Fig.~\ref{fig:Nb-vs-field}. DFT predicts this moment, due to Nb-Cr hybridization, together with a Cr moment closer to the spin-only value. The \Nb\ NMR data supported by the DFT insight confirm the microscopic spin structure suggested in Ref.~\cite{ogloblichev2021electronic}.

DFT calculations of the EFG tensor show that its principal component, $V_{zz}$, is along $\hat c$ and orthogonal to \Bhf~in ZF, establishing that the splitting of the $m=\pm1$ satellites is the quadrupole frequency of Tab.~\ref{tab:hfq} \cite{ogloblichev2017magnetic}.

 A prior study \cite{hall2022comparative} indicates that $\mathbf H \perp \hat c$ generates the expected  $2\pi$ domain walls corresponding to magnetic solitons (CSL), giving way to the FFM phase around $\mu_0 H  = 40$ mT. In our data this is a tiny field compared to the substantial \bhf = 25 T, barely showing up as a flattening of the shifts (Fig.~\ref{fig:Cr_NMR} (b) towards $H=0$. 
 
The transition of the CHM to the CCP by applying the field along the chiral axis ($\mathbf H\parallel \hat c $) is neatly witnessed by the fit of Fig. \ref{fig:conical} (b), and the simultaneous agreement of the extracted conical angle $\theta(H)$ of panel (c) with the mean-field theory. \cite{kousaka2009chiral} This provides a precise measurement of  the boundary to the FFM phase, $\mu_0 H_c = 1.35(1)$ T, for our sample. The fit deviation of the point at $\mu_0H = 40$ mT on the left of the vertical dashed line in Fig.~\ref{fig:Cr_NMR} which coincides within errorbars with the ZF value, demonstrates the presence of a pinned CHM phase at low fields in our crystal, i.e.~of a small coercive field required to initiate the spin canting. The very fine agreement of the canted model also establishes the model of the joint Eqs.~\ref{eq:frequencyc} and \ref{eq:meanfieldc} as a strong validation of the presence of the conical phase. 

The first-order nature of the transition at \TN~is very clearly shown by NMR, a local probe which can separate the decrease of the magnetic phase volume from that of the moment, a distinction that neither diffraction nor macroscopic techniques can easily provide. This finding is consistent with mean field predictions \cite{laliena2016} and a previous analysis of the entropy change at the transition \cite{clements2017critical} in a work that attributes the difficulty of measuring this change due to a weak first-order nature. The NMR convincing signature is a more direct assessment of the same fact.



All the above conclusions based on clean single crystal \Cr\ NMR data, are independent of SOD and demonstrate the expected behavior of a monoaxial chiral helimagnet. The \Nb~NMR spectra in a polycrystal sample with larger disorder \cite{SM} further demonstrate a sizable spin polarization at Nb, opposite to that of nn Cr ions.

In summary, we may classify the \Cr~NMR data from our single crystal and polycrystal samples as a interesting textbook case for an ideal monoaxial chiral helimagnet.

\subsection{\MNS}
\label{sec:discusMNS}

Our new interesting findings are on the Mn compound, whose magnetic phase diagram \cite{hall2022comparative} is still lacking a precise identification. The three component NMR spectra are assigned to a minority \MfNS~inclusion and to two structural variations of the dominant \MNS~phase.

A previous work \cite{hall2022comparative} has shown a partial occupation of the Mn $2b$ site in the very same samples used in this work. First principles simulations show that the additional Mn moment is slightly reduced (by $\sim$~2.5\%) and aligned antiparallel to its three neighbouring Mn $2c$ atoms. Its presence reflects on the neighbouring Mn atoms as well, by producing a similar reduction of their local magnetic moment, and it may be qualitatively justified by opposite signs of the on-site and transferred hyperfine couplings for the nn sites (Sec.\ref{sec:NMR-NQR}).
The reduction compares reasonably well with the experimental ratio $({^{55}\nu}_{II}-{^{55}\nu}_{I} )/\overline{{^{55}\nu}} = 0.043$, thus assigning component I to the \Mn~nuclei with one nearest neighbor (nn) defect, a Mn in the $2b$ site, and component II to the less perturbed nuclei, more removed from the defect.
A simple binomial simulation of the $2b,d$ Mn sites nn to $2c$, assuming the refined occupancies of Ref.~\cite{hall2022comparative} (0.85, 0.08 and 0.02 for Mn sites  $2c$, $2b$, and $2d$ respectively) indicates that two main environments are most probable: those with no nn defects ($p=0.73$) and those with one nn defect of either $2b$ or $2d$ type ($p=0.24$). This supports our observation of two groups of resonances in single crystal \MNS~in Fig.~\ref{fig:Mn_hllab}. 
 
This qualitative attribution is confirmed  by the behavior of the quadrupole splitting observed through the well known mechanism of spin echo oscillations shown in Fig.~\ref{fig:Mn_hllc} (c). The successful fit for  $\mathbf H \parallel \hat c$  to Eq.~\ref{eq:spinecho} is plotted in Fig~\ref{fig:Mn_qd} (a), (b) and supports the chiral conical model. The success relies on the homogeneity of the quadrupole coupling for all nuclei, despite the disorder-induced, broad magnetic inhomogeneous distribution of hyperfine fields, which hampers a better refinement of the \Mn~NMR spectrum. In addition, the first moment of the two low frequency \Mn~ components in Fig.~\ref{fig:Mn_hllc} (b) is also compatible with the same model, despite the disorder: the negative slopes agree with $^{55}\gamma$ for $\mathbf H \perp \hat c$ and a much reduced initial field dependence is observed for $^{55}\gamma$ for $\mathbf H \parallel \hat c$, like in the \Cr~case. This peculiar behavior, leading to very different onset fields for the FFM phase in the two orientations, also clearly identifies the majority phase as a chiral helical magnet, with rather larger phase boundaries to the forced ferromagnet phase, in agreement with magnetometry data \cite{SM}. 

\subsection{Conclusions}
We have shown that \CNS~ is a textbook case of a chiral monoaxial helimagnet, and we have improved the determination of its CCP to FFM phase boundary at $T=0$. This allowed us to establish a very detailed model for the NMR of a CHM, and leveraging on this we have further shown that, despite the much higher degree of intrinsic disorder, \MNS~also fully qualifies as a chiral helical magnet.

\section{Data availability}
\label{sec:data}
The original data in this work is available at the Archive Material Cloud. \cite{MC}
\\


\section{Declaration of competing interest}

The authors declare that they have no known competing
financial interests or personal relationships that could have
appeared to influence the work reported in this paper.


\section{Acknowledgments}

This work was supported by the Deutsche Forschungsgemeinschaft (DFG) within the W\"urzburg-Dresden Cluster of Excellence on Complexity and Topology in Quantum Matter -- \textit{ct.qmat} (EXC 2147, project-id 390858490) and the SFB 1143 (project-id 247310070). Work in Parma was funded by the PNRR MUR project ECS-00000033-ECOSISTER.
The computational resources were provided by the SCARF cluster of the STFC Scientific Computing Department and by the ISCRA initiative of CINECA with project IsCb6\_TRSBKS.
The work at the University of Warwick was funded by EPSRC, UK through Grants EP/T005963/1 and EP/N032128/1.  

\bibliography{references}

\end{document}